\documentclass[12pt]{article}
\usepackage{amssymb,a4wide}

\def\be{\begin{equation}}
\def\ee{\end{equation}}

\def\tb{t_{\scriptscriptstyle \rm B}}

\def\tf{t_{\scriptscriptstyle \rm F}}
\def\ti{t_{\scriptscriptstyle \rm I}}

\def\t{\widetilde}
\def\rhol{\rho_{\lambda}}

\begin{document}

\title{Brane World Cosmology Without the $Z_2$ Symmetry}

\author{
Anne-Christine Davis\footnote{A.C.Davis@damtp.cam.ac.uk},
Ian Vernon\footnote{I.R.Vernon@damtp.cam.ac.uk}, \\
\normalsize \em Department of Applied Mathematics and Theoretical Physics,\\
\normalsize \em University of Cambridge, Cambridge, CB3 9EW, UK. \\ \\
Stephen C. Davis\footnote{S.C.Davis@swansea.ac.uk} \ and
Warren B. Perkins\footnote{w.perkins@swansea.ac.uk} \\
\normalsize \em Department of Physics, University of Wales Swansea,\\
\normalsize \em Singleton Park, Swansea, SA2 8PP, Wales
}

\date{\today}

\maketitle

\begin{abstract}
The Friedmann equation for a positive tension brane situated between
two bulk spacetimes that posses the {\it same} 5D cosmological
constant, but which does not posses a $Z_2$ symmetry of the metric
itself is  derived, and the possible effects of dropping the $Z_2$
symmetry on the expansion of our Universe are examined; cosmological
constraints are discussed. The global solutions for the metric in the
infinite extra dimension case are found and comparison with the
symmetric case is made. We show that any brane world senario of this
type must revert to a $Z_2$ symmetric form at late times, and hence
rule out certain proposed scenarios.
\end{abstract}

\setcounter{page}{0}
\thispagestyle{empty}

\vfill

\begin{flushright} 
DAMTP-2000-79  

SWAT/264A
\end{flushright}

\vfill

\section{Introduction}

Recently there has been considerable interest in the novel suggestion
that we live in a Universe that possesses more than four
dimensions. The standard model fields  
are assumed to be confined to a hypersurface (or 3-brane)
embedded in this higher dimensional space, in contrast the gravitational
fields propagate through the whole of
spacetime~\cite{brane1,brane2,Sundrum,Rand1}. 
In order for this to be a
phenomenologically relevant model of our universe, 
standard 4D gravity must be recoverd on our
brane. There are various ways to do this, the most obvious being to
assume that the extra dimensions tranverse to our brane are compact.
In this case gravity can be recovered on scales larger than the size
of the extra dimensions~\cite{brane2}. This is different from earlier 
proposals since the restrictions on the size of the 
extra dimensions from particle
physics experiments no longer apply, as the standard model fields 
are confined to the brane. The extra dimensions only have to
be smaller than the scale on which gravity experiments have probed,
currently of order 1mm.
Another way to recover 4D gravity at large distances is 
to embed a positive tension 3-brane into an
$AdS_5$ bulk~\cite{Rand1,Rand2}. In this scenario 4D gravity is obtained
at scales larger than the $AdS$ radius. Randall and Sundrum showed
that this could produce sensible gravity even if the extra dimension
was not compact. 

The cosmology of these extra dimension scenarios has been investigated
and the Friedmann equation derived and shown to contain 
important deviations from the usual 4-dimensional
case~\cite{Bine1}. Some inflationary models have been
investigated~\cite{wands}, as have brane world phase transitions, topological
defects and baryogenesis~\cite{paper2}.
Most brane world scenarios that involve one extra dimension 
assume a $Z_2$ symmetry about our brane, 
motivated by a model derived from M-Theory proposed by Horava and 
Witten~\cite{Hora}. However, many recent papers examine
models that are not directly derived from M-Theory: for example there
has been much interest in the one infinite extra dimension proposal.
The motivation for maintaining the $Z_2$ symmetry is seemingly no
longer adequate and it is therefore interesting to analyse a brane
world model without this symmetry and to assess its phenomenological
implications. There have been some 
multi-brane scenarios suggested, which involve branes that although lying
between two bulk space times with the same cosmological constant, 
{\it do not possess 
a $Z_2$ symmetry of the metric itself}~\cite{Ross}. 
This approach generates an altered Friedmann equation as well as 
giving different bulk solutions. 
The cosmological
solutions have not been analysed and it is 
therefore interesting to
entertain the possibility that we live on such a brane and to
determine whether it possesses a sensible cosmology. A different
approach has been taken by~\cite{Nat,Wass} where they have looked at
the effect of having different cosmological constants either side of
the brane. Our approach is not equivalent to theirs, as will be
discussed in more detail below.
 
In Section 2 of this paper we utilise the 5D Einstein's equations to 
derive the Friedmann 
equation for such a non-$Z_2$ symmetric 3-brane and we then
investigate the 
associated 
cosmological consequences of such a model. We show that the Friedmann
equation acquires an extra term when there is no $Z_2$ symmetry, which
can give rise to a period of expansion on the brane, and
consider constraints on such a term from nucleosynthesis. These
constraints apply to any scenario in which we reside on a 3-brane
without the $Z_2$ symmetry, and demonstrate that scenarios such
as~\cite{lykken} are physically unrealistic. 
In Section 3 we restrict ourselves to the
infinite fifth dimensional case and solve Einstein's equations in the 
bulk in order to generate the 
global solutions, examining the difference between them and the
symmetric solutions and also demonstrating that they remain
well-defined. Our conclusions are summarised in Section 4.

\section{The Friedmann Equation with no $Z_2$ Symmetry}

In this section we examine a cosmologically realistic positive tension
3-brane in 5 dimensions. We assume that the 5D cosmological constants
either side of the brane are identical and then derive the general solution to
Einstein's equations without assuming that the metric across the brane
is $Z_2$ symmetric. For 
the single brane case, this amounts to having non-$Z_2$
symmetric initial conditions with a symmetric bulk, however many of
the results also apply to multi-brane scenarios such as~\cite{Ross}.

Since we are interested in 
cosmological solutions, we take a metric of the form:
\begin{equation}\label{met}
  ds^2 = -n^2(t,y) dt^2 + a^2(t,y) \gamma_{ij} dx^i dx^j + b^2(t,y) dy^2,
\end{equation}
where $y$ is the coordinate of the fifth dimension and we adopt a 
brane-based approach where the brane is the hypersurface defined by
$y=0$. Here $\gamma_{ij}$ is a maximally symmetric 3-dimensional
metric with $k=-1,0,1$ parameterising the spatial curvature. We assume 
immediately that $b^2(t,y)$ is not a function of time
and therefore $y$ can be scaled so that $b(y) = 1$. The metric is found 
by solving the 5D Einstein's equations,

\begin{equation}
G_{AB} = \kappa^2 T_{AB},
\end{equation}
where we define $\kappa^2 = 1/\t{M}_5^3$ where $\t{M}_5$ is the 
fundamental (reduced) 5D Planck Mass. The stress-energy-momentum 
tensor can be written as,

\begin{equation}
 T^A{ }_B = T^A{ }_B|_\mathit{brane} + T^A{ }_B|_\mathit{bulk}.
\end{equation}
Again with cosmology in mind, we assume a homogeneous and isotropic 
geometry in the brane and this makes it possible to write the first term as:

\begin{equation}
 T^A{ }_B|_\mathit{brane} = \delta(y) {\rm diag}(-\rho_b,p_b,p_b,p_b,0).
\end{equation}
The second term, which describes the same negative bulk cosmological 
constant $\rho_B$ either side of the brane, is of
the form,

\begin{equation}
 T^A{ }_B|_\mathit{bulk} = {\rm diag}(-\rho_B,-\rho_B,-\rho_B,-\rho_B,-\rho_B).
\end{equation}
Note that here we have taken an alternative approach
from~\cite{Nat,Wass} who assume different cosmological constants
either side of the brane. Our lack of $Z_2$ symmetry enters when we
consider the solution of the metric itself. As is shown
in~\cite{Bine2}, for the setup defined above, any set of 
functions $a(t,y)$ and $n(t,y)$
satisfying, 

\begin{equation} \label{bulk_eqtn}
\left( 
      \frac{\dot{a}}{n a}
\right)^2  \; = \; 
\frac{1}{6}\kappa^2\rho_{B}  \; + \; 
\left( 
      \frac{a'}{a}
\right)^2 \; - \; \frac{k}{a^2} \; + \; \frac{\mathcal{C}}{a^4},
\end{equation}
together with $G_{05} = 0$, will be solutions to all the Einstein's
equations locally in the bulk. The last term in (\ref{bulk_eqtn})
comes from the electric part of the 5D Weyl tensor (where
$\mathcal{C}$ is some constant) and its existence and possible
problematic effects will be discussed below. Now $G_{05} = 0$ is satisfied if

\begin{equation} \label{defn_n}
n(t,y) \;=\; \frac{\dot{a}(t,y)}{\dot{a}(t,0)}.
\end{equation}
Here, we have normalised $n(t,y)$ so that $n(t,0) = 1$. In order to 
obtain the Friedmann Equation on the brane, we 
evaluate (\ref{bulk_eqtn}) at $y=0$. This is easily done except for the
$(a'/a)^2$ term. To evaluate this we need the junction conditions,

\begin{eqnarray} \label{junc_cond}
 \frac{[a']}{a_0} & = & -\frac{\kappa^2}{3} \rho_b,  \\
\label{junc_cond2} 
 \frac{[n']}{n_0} & = &  \frac{\kappa^2}{3}(3p_b +2\rho_b),
\end{eqnarray}
where $[Q] = Q(0^+) - Q(0^-)$ and $Q(0) = Q_0$. Now instead of 
assuming the $Z_2$ symmetry $y \leftrightarrow -y$ on the metric
itself which
would give $a'(0^+) = - a'(0^-)$ and therefore $[a'] = 2a'(0^+)$, we
write,

\begin{equation} \label{defn_d}
a'(0^+) = -a'(0^-) + d(t). 
\end{equation}
where $d(t)$ is some function of time only and has yet to be determined.
$d(t)$ represents the asymmetry of the metric across the brane and the
fact that it is not identically zero for all time is what is referred
to as the brane being `non-$Z_2$ symmetric' in the rest of this paper. 
Now in order to determine $d(t)$ we use (\ref{junc_cond}) which gives,

\begin{eqnarray} \label{new_derv}
a'(0^+) & = & -\frac{\kappa^2}{6} \rho_b a_0 + \frac{d(t)}{2},\\
\label{new_derv2}
a'(0^-) & = &  \frac{\kappa^2}{6} \rho_b a_0 + \frac{d(t)}{2}.
\end{eqnarray}
Assuming that $a'^2(0) = (a'^2(0^+) + a'^2(0^-))/2$ results in,

\begin{equation}
\frac{a'^2_0}{a^2_0} \; = \; \frac{\kappa^4}{36} \rho^2_b +
\frac{d^2(t)}{4 a_0^2}.
\end{equation}
To find an expression for $d(t)$ we take the jump of the (5,5)
component of Einstein's equations as is done in~\cite{Bine1},

\begin{equation}
\frac{\bar{a}'}{a_0} p_b \; = \; \frac{1}{3} \rho_b
\frac{\bar{n}'}{n_0},
\end{equation}
where $\bar{Q} = (Q(0^+) + Q(0^-))/2$. Replacing $a$ and $n$ using
equations (\ref{defn_d}) and (\ref{defn_n}), shows that,

\begin{equation} \label{ddot}
\dot{d(t)} \; = \; \frac{3p_b \dot{a_0}}{\rho_b a_0} \: d(t).
\end{equation}
The energy conservation equation is derived directly from the junction
conditions (\ref{junc_cond}) and (\ref{junc_cond2}) as shown by~\cite{Bine1},

\begin{equation} \label{energy}
\dot{\rho_b} \; = \; -3 (\rho_b + p_b) \frac{\dot{a}_0}{a_0}.
\end{equation}
By using this to solve the differential equation (\ref{ddot}), 
it gives us the desired expression for $d(t)$,

\begin{equation}\label{defnd}
d(t) \; = \; \frac{2F}{\rho_b a_0^3},
\end{equation}
where $F$ is an integration constant which, when non-zero dictates to
what extent the $Z_2$ symmetry is broken. Combining this with
equation (\ref{bulk_eqtn}) and doing the usual replacements (first
found by~\cite{csaki}) to obtain the 
standard $H^2 \propto \rho$ relation at late times: 
$\rho_b = \rho + \rho_{\lambda}$
and $\kappa^2 \rho_B/6 + \kappa^4 \rho^2_{\lambda}/36 = 0$, (where
$\rhol$ is the brane tension and $\rho$ is the physical brane energy
density) results in
the Friedmann equation for a brane without the $Z_2$ symmetry,

\begin{equation} \label{Fried_eqtn}
\left( 
      \frac{\dot{a}_0}{a_0}
\right)^2 = 
\frac{\kappa^4 \rho_{\lambda}}{18} \rho  \; + \; 
\frac{\kappa^4}{36}\rho^2 \; - \; \frac{k}{a_0^2} \; + \; 
\frac{\mathcal{C}}{a_0^4} \; + \; \frac{F^2}{(\rho + \rho_{\lambda})^2 a_0^8}.
\end{equation}
So the absence of the $Z_2$ symmetry gives rise to an extra term in
the Friedmann equation. For a radiation
dominated Universe where $\rho = \gamma/a_0^4$, the extra term behaves
as $F^2/\gamma^2$ as $\rho \rightarrow \infty$ and $(F\rho/\gamma
\rho_{\lambda})^2$ as $\rho \rightarrow 0$. Note that equation 
(\ref{Fried_eqtn}) is the same as that found by~\cite{Kraus}, but
differs from the Friedmann equations found
by~\cite{Nat,Wass}. Therefore all the following analysis and results
of this section also apply to the setup described in~\cite{Kraus}. In 
order to obtain standard cosmology at late times we need to make
the identification,

\begin{equation}
\frac{\kappa^4 \rho_{\lambda}}{18} = \frac{8 \pi G_4}{3} = 
\frac{1}{3 \t{M}_4^2},
\end{equation}
 where we have used the reduced 4D Planck mass defined by $\t{M}_4^2 =
M_4^2/8\pi$ and will use the 5D reduced Planck mass defined by $\t{M}_5^3 =
M_5^3/8\pi$. This implies that the brane tension,
$\rho_{\lambda} = 6\t{M}_5^6/\t{M}_4^2$. 
Using this, the
fact that $\kappa^2= 1/\t{M}_5^3$ and also defining the dimensionless
constants $f = F\t{M}_4^2/\gamma \t{M}_5^3$, 
$c = 3 \t{M}_4^2 \mathcal{C}/\gamma$ and assuming $k=0$, allows us to write
(\ref{Fried_eqtn}) in terms of $f$, $c$, $\t{M}_4$, $\t{M}_5$ (and 
$\rho_{\lambda}$),

\begin{equation} \label{Fried_eqtn2}
\left({\dot{a}_0 \over a_0}\right)^2 = 
 {1+c \over 3\t{M}_4^2} \rho 
+ {1 \over (6\t{M}_5^3)^2} \left[ \rho^2 + {f^2 \rhol^2 \rho^2 \over 
 \left(\rho + \rhol \right)^2}
\right].
\end{equation}

This shows that the expansion of the universe is initially dominated by
the $\rho^2$ term, while at late times the standard cosmology phase 
with the usual $H^2 \propto \rho$ behaviour is obtained. 
If $f < \sqrt{8(1+c)}$ the third, `$f$'-term of (\ref{Fried_eqtn2}) is always less
significant than the other terms, and the resulting cosmology is
similar to a brane cosmology with a $Z_2$ symmetry. 

If $f>3+2c$ there will be a period between the $\rho^2$ and $\rho$ driven
phases when the `$f$'-term is dominant. It is reasonable to suggest
that $f$ is significantly greater than $c$, since as will be
demonstrated later in this section, nucleosynthesis constrains $c$ to
be $\ll 0.2$. Now in order to get an explicit solution for
$a_0(t)$ so that we can ascertain the effect of the new $f$ term, we
are forced to make several approximations. First we use,

\be
{\rho^2 \rhol^2\over (\rho + \rhol)^2} = \left\{ \begin{array}{cc}
\rho^2 & \rho < {1 \over 4} \rhol \\
{1 \over 4}\rhol \rho  & {1 \over 4} \rhol< \rho < 4\rhol \\ 
\rhol^2  & \rho > 4\rhol \\
\end{array}\right. ,
\ee

to approximate the `$f$'-term. Now it is possible to solve for
$a_0(t)$ in equation (\ref{Fried_eqtn2}) which demonstrates that the
time dependance of $a_0(t)$ in each of the three phases takes the form:

\be\label{3phases}
a_0(t) \propto  \left\{ \begin{array}{cc}
\gamma \left( \frac{4(1+c)}{3\t{M}_4^2} t^2 +
            \frac{2\sqrt{1+f^2}}{3\t{M}_5^3} t 
       \right)^{\frac{1}{4}}    & \rho < {1 \over 4} \rhol \\
\gamma \left( \frac{16(1+c)+f^2}{12\t{M}_4^2} t^2 +
            \frac{2}{3\t{M}_5^3} t 
     \right)^{\frac{1}{4}}       & {1 \over 4} \rhol< \rho < 4\rhol \\ 
\gamma \left( \frac{(1+c)}{\rhol}\left[ 
                       \cosh{\frac{2\rhol f}{3\t{M}_5^3} t}-1
                                 \right]
              + \frac{1}{\rhol} \sinh{\frac{2\rhol f}{3\t{M}_5^3} t}
       \right)^{\frac{1}{4}}     & \rho > 4\rhol \\
\end{array}\right. ,
\ee

Evaluating the exact constants that replace the proportional signs
in (\ref{3phases}) is rather cumbersome, so in order to obtain an order of
magnitude approximation of the effects of the `f'-term we 
instead ignore all subdominant terms in each phase of the universe,
assume that $f\gg 1$ and
then solve for $a_0(t)$ again using (\ref{Fried_eqtn2}). The
resulting evolution of the universe then divides into 5 phases as
described below. In the following it is useful to define the time
scale $\tf=\t{M}_4^2/ (4\t{M}_5^3 f)$,

\begin{itemize}

\item PHASE 1  ($0 <t < \tf$): Initially (\ref{Fried_eqtn2}) is
dominated by the $\rho^2$ term. This continues until $t = \tf$ when
$\rho = f \rhol$, after which the `$f$'-term becomes dominant. Until
that happens,
\be
a_0(t) = \left[{2 \gamma \over 3 \t{M}_5^3}\right]^{1/4} t^{1/4}
\ee

\item PHASE 2a ($\tf < t < \tf + \ti$), with $\ti = \tf \ln (f/4)$:
For large $\rho$ the `$f$' term can be approximated by a
constant. During this period the universe expands exponentially, as in
inflation, 
\begin{equation}
a_0(t) = \left[{\t{M}_4^2 \gamma \over 6\t{M}_5^6 f e}\right]^{1/4} e^{H t} 
\ , \ \ H = {\t{M}_5^3 \over \t{M}_4^2} f.
\end{equation}

\item PHASE 2b  $(\ti +\tf < t < \ti + 7\tf$): For $\rho \sim \rhol$
the `f'-term starts to decrease, and is approximately proportional to
$f^2 \rhol \rho$. During this phase,
\begin{equation}
a_0(t) = \left[{\gamma f^2 \over 6 \t{M}_4^2}\right]^{1/4} (t-\ti+\tf)^{1/2}.
\end{equation}

\item PHASE 2c  $(\ti+ 7\tf < t < \tb=\ti + (3+f^2/2) \tf$): For small
$\rho$ the `f'-term is approximately $f^2 \rho^2$. It ceases to be the
dominant term in (\ref{Fried_eqtn2}) when $\rho = 2\rhol/f^2$. Until then,
\begin{equation}
a_0(t) = \left[{2 \gamma f \over 3 \t{M}_5^3}\right]^{1/4} (t-\ti-3\tf)^{1/4}.
\end{equation}

\item PHASE 3  $(t > \tb)$: Finally, at late times the $\rho$ term
dominates (\ref{Fried_eqtn2}) to give the standard cosmology,
\begin{equation}
a_0(t) = \left[{4 \gamma (1+c)\over 3\t{M}_4^2}\right]^{1/4} 
(t-\tb+f^2\tf)^{1/2}.
\end{equation}

\end{itemize}

The above solution for $a_0(t)$ presents a very different picture of
the evolution of our universe: it has the 
unconventional early
$\rho^2$ behaviour as seen in most brane world models, but now this is
broken up by a period of exponential expansion. During this extra phase the
scale factor increases by a factor of $f^{1/4}$. Like inflation this
could help to solve the flatness problem, however unlike inflation
it is not followed by reheating and so cannot help with the horizon and
monopole problems. Eventually, as expected, the standard
cosmology is obtained. The above approximate solution (equations
(23--27)), suggests that
there will be no exponential expansion unless $f \gtrsim 4+2c$. If $f$ is
very large ($f \gtrsim \t{M}_4^2/\t{M}_5^2$) then $f\rhol > \t{M}_5^4$
and Phase 1, and some of the succeeding phases will be above the 5D
Planck scale, and may not actually occur.

For $\sqrt{8(1+c)} < f < 3+2c$ there will be a short period of `$f$'-term
domination. This occurs after the time when the $\rho$-term of
(\ref{Fried_eqtn2}) starts to dominate the $\rho^2$-term. At this stage
the `$f$'-term is no longer approximately constant, and so
there is no exponential expansion. 

Nucleosynthesis provides constraints on both $c$ and $f$. If $c$ is
nonzero then it gives rise to a term of the form ${\mathcal{C}}/a_0^4$
in the Friedman equation (\ref{Fried_eqtn}) which at the time of
nucleosynthesis behaves like additional relativistic degrees of
freedom. The energy density at this time is given as 
$\rho(t_N) = g_* \frac{\pi^2}{30}T_N^4$, where $t_N$ and $T_N$ are the
time and temperature of Nucleosynthesis, and $g_*$, the number of
effective relativistic degrees of freedom is strongly 
constrained by the observed
abundances of light elements which show that any deviation from $g_*$
given by $\Delta g_*$ satisfies: $\Delta g_* < 2$. This implies the
following constraint for $\mathcal{C}$, as pointed out by \cite{Bine2},

\begin{equation}
\frac{\mathcal{C}}{a_0^4(t_N)} \;\; \ll \frac{\pi^2}{30}
\frac{\Delta g_* T_N^4}{3\t{M}_4^2}.
\end{equation}
Using the fact that 
$\rho(t_N) = \gamma/a_0^4(t_N) = g_*\frac{\pi^2}{30}T_N^4$, leads to,

\begin{equation}\label{constrC}
\frac{\mathcal{C}}{\gamma} \;\; \ll \;\; 
\frac{\Delta g_*}{3\t{M}_4^2 g_*},
\end{equation}
which implies that 

\begin{equation}\label{ccc}
c \;\; \ll \;\; \frac{\Delta g_*}{g_*} \simeq 0.2,
\end{equation}
where we have taken the standard values $g_* = 10.75$ and 
$\Delta g_* < 2$. Hence it is reasonable to suggest that $f$ could
play a cosmologically significant role but that $c$ cannot. Note that
this constraint on $c$ is independant of the approximations used to
derive equations (22) and (23--27) and comes purely from our
knowledge of the restrictions on extra relativistic degrees of freedom
at the time of nucleosynthesis. 

We can also obtain rough restrictions on $f$ in terms of $\t{M}_5$ by 
demanding
that standard cosmology is in place by the time of
nucleosynthesis. This is equivalent to requiring that phase 2 is over
well before nucleosynthesis begins, which implies that,

\begin{equation}\label{NucConstr1}
{1+c \over 3\t{M}_4^2} \rho(t_N) \gg {1 \over (6\t{M}_5^3)^2} f^2 \rho(t_N)^2.
\end{equation}
At this time the universe is radiation dominated and
$\rho(t_N)$ can be written in terms of the temperature at 
nucleosynthesis, $T_N \approx 1$MeV, and the number of effective
relativistic degrees of freedom, $g_*$,

\begin{equation}
\rho(t_N) = g_* \frac{\pi^2}{30}T_N^4.
\end{equation}
Substituting this into (\ref{NucConstr1}), assuming $c$ to be
negligible due to (\ref{ccc}) and switching to the standard
4D and 5D Planck masses, gives the following relation
between the 5D Planck mass $M_5$ and the $Z_2$ symmetry breaking
parameter $f$,

\begin{equation}\label{MfConstr}
M_5 \gg \left(\frac{g_* \pi^3}{45} T_N^4 M_4^2 f^2 \right)^{1/6}
 	    \approx 30f^{1/3} \mbox{TeV}.
\end{equation}
For the case where $f=0$, only phase 1 and 3 exist and demanding just 
that phase 3 is over well before $t_N$ gives, from
(\ref{Fried_eqtn2}), 

\begin{equation}\label{Nucfo}
{1+c \over 3\t{M}_4^2} \rho(t_N) \gg {1 \over (6\t{M}_5^3)^2} \rho(t_N)^2.
\end{equation}
which leads as above to:

\begin{equation}
M_5 \gg 30 \mbox{TeV}.
\end{equation}

This is a fairly weak bound on
$M_5$. For the infinite extra dimension scenario, experiments testing 
gravity at small distances have already
demonstrated using the corrections to Newton's gravity law calculated
by Randall and Sundrum~\cite{Rand2} that $M_5 > 10^5$TeV. This experimental
constraint however, is not applicable to compactified scenarios. 
Supposing that we do live in an infinite 5th dimension and that 
$M_5$ has a value that is just outside our experimental reach, we can
then use (\ref{MfConstr}) to constrain $f$; $f \ll 10^{11}$. 
In this case, the period of exponential expansion
(Phase 2a) would only last for $\ti = M_4^2 / (4 M_5^3 f) \ln f
\simeq 10^{7} \mbox{TeV}^{-1}$ and the scale factor would increase during this
time by a factor of around $500$.  
Increasing $M_5$ relaxes the bound on $f$ and would appear to lead to 
more inflation. However, if $M_5$ is too large then some, or all, of the
resulting expansion occurs at energies higher than the 5D Planck scale.
Consequently the maximum inflation occurs for $M_5 = 5 \times 10^6$ TeV and
$f = 10^{17}$, which leads to an expansion of only $10^4$. Unfortunately,
this is not cosmologically relevant. These order of magnitude
expansion rates due to a non-zero $f$ are dependant on the 
approximations made to generate equations (23-27), however an exact
treatment would likely lead to the same conclusion: that the
cosmological effect of $f$ is negligible. 

We have therefore derived constraints on $c$ and $f$, showing both 
that $c$ must be negligible in size: $c\ll 0.2$ and that although $f$ can be
large, its cosmological effect is insignificant in that it causes
inflation that results in an increase in $a_0(t)$ only of order $10^4$. 

Note that, from equations (11), (12), (17) and (31)
the effect of the $Z_2$ breaking term decreases with increasing time
such that the universe reverts to standard cosmology. This suggests
that brane world scenarios where the physical Universe is on a brane 
without this symmetry
are not viable after nucleosynthesis. The proposal made
in~\cite{lykken}, which described a setup that would solve the
hierarchy problem despite our physical universe existing on a positive
tension brane, is unfortunately found to be unrealistic due to this reason.

\section{Global Solutions}
After having examined the phenomenological effects resulting from the Friedmann
equation of a non-$Z_2$ symmetric brane world, we will 
now solve the 5D Einstein equations in the bulk and hence derive 
the corresponding global solutions for 
$a(t,y)$ and $n(t,y)$ for the non-$Z_2$ symmetric brane in an
infinite extra dimension. We derive
this solution for the general case first and then for a specific
cosmologically
realistic brane. To do this we can adapt the previously
known general global solution for a brane with tension $\rho_b$ and
negative 5D bulk cosmological constant $\rho_B$ to the non-$Z_2$
symmetric case. 

We know 
from the (0,0) component of Einstein's equations that the new 
non-symmetric solution will have a form similar to the symmetric 
case as derived in~\cite{Bine2};

\begin{equation}\label{newsoln}
a^2(t,y)  =  a^2_0 \left( 
                           A(t)\cosh \mu y + B(t)\sinh \mu
                           |y| + C(t)
                           \right),
\end{equation}
where $\mu=\sqrt{-2\rho_B/(3\t{M}_5^3)}$.
The requirement $a^2(t,0) = a_0^2$ trivially implies that $A(t)+C(t)
=1$ for all $t$. Now using equation (\ref{new_derv}) it can be seen that,
\begin{eqnarray} \label{defn_B}
a'(t,0^{\pm}) \; & = & \; \left\{a'_{sym}(t,0^{\pm}) \right\}\; + \; \frac{d(t)}{2}, \\
\Rightarrow \;\;\; B(t) \; \; \; & = & 
\; \; \; 
\left\{ -\frac{\rho_b}{\sqrt{-6 \t{M}_5^3\rho_B}} \right\}
\; \; \pm \; \; \frac{d(t)}{a_0 \mu}.
\end{eqnarray}
Here the $\pm$ in the expression for $B(t)$ corresponds to the
solution on either side of the brane and we use $\{\ldots\}$ to denote the
solution found in the $Z_2$ symmetric case. $C(t)$ is found from the 
differential equation for $a^2(t,y)$ which is
derived from the Einstein equations, (see~\cite{Bine2}), 
\begin{equation}
 C(t) = \frac{3\t{M}_5^3(\dot{a}_0^2 + k)}{\rho_B a_0^2}.
\end{equation}
Rewriting this using the `new' Friedmann equation (\ref{Fried_eqtn})
gives,
\begin{equation}
 C(t) = \left\{\frac{1}{2} \left(
                            1 + \frac{\rho^2_b}{6\t{M}_5^3\rho_B}
                      \right)
          +\frac{3 \t{M}_5^3\mathcal{C}}{\rho_B a_0^4}\right\}
          +\frac{3\t{M}_5^3F^2}{\rho_B \rho^2_b a_0^8},
\end{equation}
and therefore $A(t)$ is trivially given by,
\begin{equation}
A(t)  =  \left\{\frac{1}{2} \left(
                            1 - \frac{\rho^2_b}{6\t{M}_5^3\rho_B}
                      \right)
          -\frac{3 \t{M}_5^3\mathcal{C}}{\rho_B a_0^4}\right\}
          -\frac{3\t{M}_5^3F^2}{\rho_B \rho^2_b a_0^8}.
\end{equation}
Using (\ref{defn_B}) and (\ref{defnd}) leads to,
\begin{equation}
 B(t)  = \left\{ -\frac{\rho_b}{\sqrt{-6\t{M}_5^3\rho_B}}\right\}
 \pm
                            \sqrt{\frac{6\t{M}_5^3}{-\rho_B}} 
                            \frac{F}{\rho_b a_0^4} .
\end{equation}
Again the $\pm$ signs in the expression for $B(t)$ give the two different
solutions on either side of the brane. The solution for $n(t,y)$ in
the non-symmetric case is found from the above solution for $a(t,y)$ by
using equation (\ref{defn_n}) as before. It is easily seen that setting 
$F$ to zero in the above solutions recovers the $Z_2$ symmetric situation.

We are interested in these solutions for a cosmologically realistic
brane, so we make the same substitutions as were made to generate
(\ref{Fried_eqtn}) and also assume a radiation dominated Universe. 
Setting $\rho = \gamma/a_0^4$ where $\gamma$ is a constant, 
we obtain expressions for $A(t)$, $B(t)$ and $C(t)$ corresponding to a
brane with a viable cosmology,
\begin{eqnarray} \label{our_soln}
  A(t) & = & 1 + \chi + \frac{1}{2}\chi^2 +
              c \chi +
             \frac{f^2 \chi^2}{2\left(1+\chi \right)^2}, \\
  B(t) & = & -(1+\chi) \pm   \frac{f \chi}{(1+\chi)}, \\
  C(t) & = & -\chi - \frac{1}{2}\chi^2 -
             c \chi -
             \frac{f^2 \chi^2}{2\left(1+\chi \right)^2}.
\end{eqnarray}
Where we have defined the dimensionless variable 
$\chi = \t{M}_4^2 \rho/6\t{M}_5^6$. Equation (\ref{our_soln}) shows
that the metric off the brane is greatly altered by a large value of
$f$, in fact the `f'-terms dominate during the period when 
$f \gtrsim \chi \gtrsim 1/f$. During this time, $a(t,y)$ vanishes on
the side of the brane corresponding to the negative sign in equation
(38), at position $y_0$ which is given by:

\begin{equation}
y_0 \simeq \frac{1}{\mu} \ln\left[
                   \frac{(f+2)\chi +2}{(f-2)\chi -2} \right],
\end{equation}
and $n(t,y)$ will only become small on the opposite side at position 
$y_0$ given by:

\begin{equation}
y_0 \simeq \frac{1}{\mu} \left|
              \frac{4(\chi^2-1)}{f\chi(\chi-3)}
                          \right|,
\end{equation}
although the exact position of the zeros will depend upon the
negleglected terms in (\ref{our_soln}).
After this time, $a(t,y)$ no longer vanishes, however $n(t,y)$
vanishes roughly as for the symmetric case where $y_0$ is now given by:

\begin{equation}
y_0 \simeq \frac{1}{\mu} \ln\left[ 1 + \frac{2\chi + 
                                  \sqrt{2\chi(2\chi+c)}}
                                        {\chi(3\chi + 2c)}
\right].
\end{equation}
Calculation of the Ricci tensor and scalar show that they are both
finite at the points where $a(t,y)$ and $n(t,y)$ vanish, which implies
that these points just correspond to
coordinate singularities. A similar result has been
obtained in~\cite{muko} for the symmetric case. Since there are no 
obvious problems with the
global solutions, the only restrictions on $f$ and $c$ are those
obtained in the previous section, which are applicable to all
scenarios, not just the infinite extra dimensional one considered
here. 

We have seen how in a cosmological context, the various new features
that are presented by a non-$Z_2$ symmetric brane world as opposed to
an ordinary 4D Universe, are not necessarily dangerous and providing
that the basic parameters $M_5$, $c$ and $f$ satisfy some fairly
relaxed constraints, then the phenomenological predictions of the
scenario proposed in this paper are still in agreement with current
experimental observation. This has been shown to be true, however only for the
cosmological case. While it is comforting to know that the Friedmann
equation for a homogeneous, isotropic brane world gives sensible late
time cosmology, the extra features such as the electric part of the 5D
Weyl tensor and the extra degrees of freedom in the metric due to
dropping the $Z_2$ symmetry condition, could cause major problems when
pertubations to the background metric are considered. Although several
papers have been produced addressing the issue 
(see~\cite{pert1,pert2,pert3,pert4}), the complete analysis remains to be
done. Only a fully rigorous
treatment would be able to show whether or not the Weyl tensor or the lack of
a $Z_2$ symmetry would lead to pertubations growing in a disasterous
manner and whether a possible choice of initial conditions could help
prevent this.

\section{Conclusions} 

We have derived 
the Friedmann equation and bulk solutions in a brane world scenario where
there is no $Z_2$ symmetry of the metric across the
brane. The bulk solutions were shown to be well behaved.
Relaxing the $Z_2$ symmetry introduces an extra term that behaves as
an effective cosmological constant at early times. 
Approximate solutions of the Friedmann equation on the brane were
found with this extra term.
If the asymmetry is
sufficient this will introduce a period of exponential 
expansion during the early {\it brany} evolution.

Demanding that standard cosmology is in place by
the time of nucleosynthesis constrains the $Z_2$ symmetry
breaking parameter.
This limits the amount of exponential 
expansion of the scale factor at early times to be of order $10^4$,
which is not cosmologically significant. 
Thus this would only be a small mitigating factor in the flatness problem.

We also note that 
the effects of the $Z_2$ breaking term decrease 
with time. This is essential to ensure that the standard
cosmology is recovered at late times. This suggests that the scenarios 
without this symmetry at
late times, such as~\cite{lykken} are not viable.

\section*{Acknowledgements}
We would like to thank Pierre Binetruy, Tim Hollowood, Nathalie Deruelle
and David Wands for discussions and PPARC for financial support.

\def\Journal#1#2#3#4{{#1}{\bf #2}, #3 (#4)}
\def\npb{Nucl.\ Phys.\ {\bf B}}
\def\pl{Phys.\ Lett.\ }
\def\prl{Phys.\ Rev.\ Lett.\ } 
\def\prd{Phys.\ Rev.\ D }
\def\jpa{J.\ Phys.\ {\bf A}}


\end{document}